\documentclass[num-refs]{wiley-article}
\usepackage{siunitx}
\usepackage{graphics} % for pdf, bitmapped graphics files
\usepackage{times} % assumes new font selection scheme installed
\usepackage{amsmath} % assumes amsmath package installed
\usepackage{amssymb}  % assumes amsmath package installed
\usepackage{color}
\usepackage[export]{adjustbox}
\usepackage{subfigure}
\usepackage[T1]{fontenc}
\usepackage{ae,aecompl}
\usepackage{soul}
\usepackage{subfigure}

\papertype{Original Article}
\paperfield{Journal Section}

\title{Gaussian Random Number Generator: {\toreferee Implemented in FPGA for Quantum Key Distribution}}

\abbrevs{FPGA: Field Programmable Gate Array; QKD: Quantum Key Distribution; GRNs: Gaussian Random Number; URNs: Unique Random Numbers; LSFR: Linear Feedback Shift Register; MSRG: Multi-return Shift Register Generator.}

\author[1]{Yue Hu}
\author[1]{Yan Wu}
\author[1]{Yi Chen}
\author[1]{Guo Chun Wan}
\author[1]{Mei Song Tong}

\affil[1]{College of Electronics and Information Engineering, Tongji University, Shanghai, 201804, China}

\corraddress{Yue Hu, College of Electronics and Information Engineering, Tongji University, Shanghai, 201804, China}
\corremail{tjhuyue@tongji.edu.cn}

\presentadd[\authfn{2}]{College of Electronics and Information Engineering, Tongji University, Shanghai, 201804, China}

\fundinginfo{ National Undergraduate Innovation Program 2016,\\College of Electronics and Information Engineering, Tongji University}

\runningauthor{Y. Hu et al.}

\begin{document}

\maketitle

\begin{abstract}
{\toreferee Quantum Key Distribution is the process of using quantum communication to establish a shared key between two parties. It has been demonstrated the unconditional security and effective communication of quantum communication system can be guaranteed by an excellent Gaussian random number generator with high speed and an extended random period. In this paper, we propose to construct the Gaussian random number generator by using Field-Programmable Gate Array (FPGA) which is able to process large data in high speed. We also compare three algorithms of GRN generation: Box-Muller algorithm, polarization decision algorithm, and central limit algorithm. We demonstrate that the polarization decision algorithm implemented in FPGA requires less computing resources and also produces a high-quality Gaussian random number, through the null hypothesis test. }

% Please include a maximum of seven keywords
\end{abstract}

\section{INTRODUCTION}

 As the modern physics suggests, quantum mechanics has a significant influence in engineering, such as quantum computing \cite{gruska1999quantum,deutsch1985quantum} and quantum tunneling electronic devices\cite{bouchiat1998quantum}. At the same, It also opens a new perspective in offering interesting new protocols in the intersection between computer science \cite{gruska1999quantum}, information theory\cite{bennett1998quantum}, and quantum cryptographic key (QCK) \cite{Herrerocollantes2017Quantum,braunstein2005quantum,cerf2001quantum}. The most well known and developed application of quantum cryptography is quantum key distribution (QKD)\cite{shor2000simple,scarani2009security}, which is the process of using quantum communication to establish a shared key between two parties without a third party learning anything about that key, even if all communication is being eavesdropped\cite{shor2000simple}. However, the absolute security of QCK is guaranteed by naturally quantum mechanical properties
 \cite{zeng2010quantum}, i.e., Heisenberg's uncertainty principle\cite{busch2007heisenberg}. In practical, the unconditional security is achieved by using quantum vacuum fluctuations
 \cite{zeng2010quantum,milonni2013quantum}, the phase noise of lasers\cite{lo2014secure}, and amplified spontaneous emission\cite{chapuran2009optical,2018MNRAS.480.1333H}. Vacuum fluctuations method is based on shot noise measurement\cite{Gabriel2010A}. The phase noise method use laser phase noise \cite{Jofre2011True} and the spontaneous emission method also utilizes fluctuations in ASE noise\cite{Williams2010Fast}\cite{Argyris2012Sub}. Recently there is a large number of proposals, experiments, improvements and exciting theoretical results in randomness extraction and randomness certification for dealing with QCK. Unfortunately, direct fluctuation measurement is not technologically feasible for optical signals, especially considering the problems from sampling and digitization.

{\toreferee However it is possible to obtain a pseudo-random quantum signal by a set of high-quality Gaussian random numbers \cite{Thomas2009A}. This requires the period of Gaussian random numbers (GRNs) is long enough for guaranteeing the absolute security of QCK. On the other side, the speed of generating is also critical for effective communication between two parties\cite{laudenbach2017continuous}.} Hence, an adequate Gaussian random source is economical and vital for continuous-variable quantum cryptography communication \citep{Wang2006}. {\toreferee Whereas, }the study of the Gaussian random numbers generator in continuous-variable QCK system is much complicated.
{\toreferee Some excellent surveys of the Gaussian random number generators (GRNGs) from the algorithmic perspective exist in the published literature of \citet{thomas2007gaussian}. \citet{thomas2007gaussian} compared their computational requirements and examined the quality. In this work, we choose three conventional algorithms for generating GRN: Box-Muller algorithm\cite{box1958,thomas2007gaussian}, polarization decision algorithm\cite{Bell:1968:ANR:363397.363547,thomas2007gaussian}, and central limit algorithm\cite{Knuth:1997:ACP:270146,thomas2007gaussian}. The Box-Muller transform is one of the earliest exact transformation methods\cite{thomas2007gaussian}. It produces a pair of Gaussian random numbers from a couple of uniform numbers. The polar method is an exact method related to the Box-Muller transform and has a closely related two-dimensional graphical interpretation, but uses a different approach to get the 2D Gaussian distribution\cite{Bell:1968:ANR:363397.363547,thomas2007gaussian}. The probability density function describing the sum of multiple uniform random numbers is obtained by convolving the constituent probability density function. Thus, by the central limit theorem, the probability density function of the sum of K uniform random numbers over the range (0, 1) will approximate a Gaussian distribution. Furthermore \citet{Malik:2016:GRN:2988524.2980052}  provides a potential  capsulization of hardware Gaussian random number generator architectures.}
In this work, we analyzed, and {\toreferee compare these three algorithms} of generating Gaussian Random Number for a continuous-variable quantum cryptography communication system. {\toreferee More importantly, we choose FPGA as hardware to construct GRNG architectures using the Box-Muller algorithm, polarization decision algorithm, and central limit algorithm. Generally, CPU executes instructions with extremely high speed. However, for a typical instruction (i.e., instruction for GRN and QCK), it is possible to achieve a higher speed through an appropriately designed architectures\cite{kestur2010blas,sidhu2001fast,underwood2004fpgas}. }

In what follows, in Section \ref{sec:2}, we briefly describe {\toreferee how to implement Gaussian random number in QCK system} by the reverse reconciliation protocol. Then in Section \ref{sec:3}, we introduce how to achieve the Box-Muller algorithm {\toreferee, polarization decision algorithm, and central limit algorithm}. In Section \ref{sec:4}, we present the designed {\toreferee architectures using FPGA circuit} and the implement of the Box-Muller algorithm{\toreferee, polarization decision algorithm, and central limit algorithm}. In Section \ref{sec:5}, we describe the hardware structure to achieve our design. In Section \ref{sec:6} we estimate the accuracy and quality of generated GRNs {\toreferee as well compare these three algorithms}. In Section \ref{sec:7}, we give the  discussion about our design of GRNs generator to be implemented in a continuous-variable QCK system. In Section \ref{sec:8}, we give our conclusions.

\section{Application of Gaussian Random Number in Quantum Key Distribution System}
\label{sec:2}
In the field of quantum key distribution, randomness is an essential requirement. Even if the communication channel is eavesdropped by others, communication on this channel is still safe with the randomness.
{\toreferee Here we introduce a coherent-state QKD protocol, whose security relies on the distribution of a Gaussian key obtained by continuously modulating the phase and amplitude of GRNs\cite{cerf2001quantum} Alice's (emitter) side, and subsequently detected at Bob's (receiver) side\cite{grosshans2003quantum}.

The protocol runs as follows\cite{laudenbach2017continuous}, Alice prepares displaced coherent states with quadrature components $q$ and $p$ that are realizations of two independent and identically distributed random variables $Q$ and $P$.
The random variables $Q$ and $P$ obey the same zero-centered normal distribution:
\begin{equation}
Q\sim P\sim N(0, V)
\end{equation}
where $V$ is referred to as variance. The displaced coherent states$|\alpha_{1}\rangle,...,|\alpha_{j}\rangle,...,|\alpha_{n}\rangle $ are expressed as:
\begin{equation}
|\alpha_{j}\rangle=|q_{j}+ip_{j}\rangle
\end{equation}
The coherent states obey the usual eigenvalue equation:
\begin{equation}
\frac{1}{2}(\hat{q}+i\hat{q})|\alpha_{j}\rangle=(q+iq)|\alpha_{j}\rangle
\end{equation}
where$\hat{q}$ and $\hat{q}$ are the quadrature operators, defined in the framework of shot-noise units\cite{madsen2012continuous}. After preparation of each coherent state, Alice transmits $|\alpha_{j}\rangle$ to Bob through a Gaussian quantum channel. Bob uses heterodyne detection to measure the eigenvalue of either one or both of the quadrature operators.}
In the last step, Bob sends the correct data to Alice and then Alice corrects her own message which have the same values as Bob.

As what we have illustrated above, to get useful key elements, at the first step Gaussian random numbers are needed to modulate the amplitude and phase information.
On the other hand, the noise that the Gaussian random source introduces to the transmission system degrades the security. {\toreferee Thus, a high-quality Gaussian random number generator plays a significant role in the QKD system.} Secondly, the high transmission speed is another advantage of quantum communication over a classical communication system. To improve the transmission speed further, a faster GRNs generator is required.

In what follows, we proposed a Gaussian random source with high output speed and low quantizing noise to efficiently generate a secure sequence of quantum key.

\section{Algorithm for Generating Gaussian Random Number}
\label{sec:3}
%\subsection{Comparison between Different Algorithms}
%After comparing several widely-used algorithms for getting Gaussian Random Numbers, finally, we choose the Box-Muller Algorithm.

%\begin{table}[htbp]
% \centering
%\label{Tab.1}
% \begin{tabular}{|c|c|c|c|c|c|c|c|}
%  \hline
% \tabincell{c}{Method} & \tabincell{c}{Uniform \\Random \\Number} & Adder & Subtracter & Divider & CMP & Ln(x) & Sqrt(x)\\
%  \hline
% \tabincell{c}{Box-\\Muller \\ Algorithm}  & 2 & 0 & 2 & 1 & 0 & 1 & 1\\
%\hline
% \tabincell{c}{Polarization \\ Decision \\ Algorithm}  & 1.27 & 1.91 & 3.27 & 1 & 1.27 & 1 & 1\\
%\hline
% \tabincell{c}{Central \\ Limit \\ Algorithm} & 2 & 0 & 2 & 1 & 0 & 1 & 1\\
%\hline
% \end{tabular}
%\caption{Resources Utilization of Different Methods. The table compares the usage of adder, subtracter, divider and other relevant operation modules in several different algorithms.}
%\end{table}

% \textbf{Table.1} gives these three different algorithms\cite{Zhang2014Algorithm} for generating Gaussian random numbers. Comparing with the Box-Muller Algorithm
%proposed in this paper, the secondly efficient method is Polarization Decision Algorithm (PDA) ,but PDA uses more adders, dividers and comparators which will make the calculation more difficult and the operation slower.

%Taking all of these factors into consideration, we find that Box-Muller is  more efficient.

\subsection{Box-Muller Algorithm}
{\toreferee The Box-Muller transform proposed by \citet{box1958}}, is one of the precise algorithms for getting the Gaussian random number. {\toreferee The Box-Muller Algorithm is based on a property of a two-dimensional Cartesian system, assuming X and Y coordinates are described by two independent and normally distributed random variables (i.e. $f_{X}(x)=\frac{1}{\sqrt{2\pi}\delta }e^{-\frac{x^2}{2\delta ^2}}$ and $f_{Y}(y)=\frac{1}{\sqrt{2\pi}\delta }e^{-\frac{y^2}{2\delta ^2}}$). If transform X and Y to the corresponding polar coordinates variables $r^{2}$ and $\theta$, the random variables $r^{2}$ and $\theta$ are also independent and can be expressed as: $f_R(r)=\int_0^{2\pi} \frac{r}{{2\pi}\delta ^2}e^{-\frac{r^2}{2\delta ^2}} \,d\theta$, $0\leqslant r \leqslant \infty $ and $f_\Theta(\theta)=\int_0^{\infty} \frac{r}{{2\pi}\delta ^2}e^{-\frac{r^2}{2\delta ^2}} \,dr$, $0\leqslant \theta \leqslant 2\pi$,} where \textbf{R} obeys the Rayleigh distribution and $\Theta$ obeys the uniform distribution, so {\toreferee their joint probability density is} $f_{R\Theta}(r,\theta)=f_{R}(r)\times f_{\Theta}(\theta)$ which is also statistically independent. Then the corresponding distribution functions {\toreferee of \textbf{R} and $\Theta$} are:

\begin{equation}
\begin{aligned}
F_R(r)=\int_0^{r} \frac{r'}{\delta ^2}e^{-\frac{r'^2}{2\delta ^2}} \,dr' \\
F_\Theta(\theta)=\int_0^{\theta} \frac{1}{2\pi}\,d\theta' =\frac{\theta}{2\pi}\\
\end{aligned}
\end{equation}

{\toreferee Fortunately, the} distribution functions $F_{R}(r)$ $F_{\Theta}(\theta)$ {\toreferee is in} closed form. Hence the Gaussian random variables can be generated by the inverse transformation method\cite{thomas2007gaussian}. {\toreferee Since the $F_{R}(r) \in [0,1]$, as well as $F_{\Theta}(\theta)$, actually} the  Gaussian random variables \textbf{X} and \textbf{Y} in Cartesian coordinates {\toreferee can be obtained through a transformation of two sets of uniformly and independently distributed random numbers}.

{\toreferee In practice, the Box-Muller algorithm samples two uniform distribution on the interval (0, 1) and then mapping them to two standards, Gaussian distributed samples with zero expectation and unit variance. The algorithm is implemented as follow}:

\begin{enumerate}
\item Generate a pair of uniformly and independently distributed random numbers between the interval (0,1), denoted as $U_{1}$ and $U_{2}$ respectively.

\item Mapping the random point to the Cartesian coordinate axis through the transformation:

\begin{equation}
\begin{aligned}
\alpha(U_1,U_2)=\sqrt{-2ln(U_1)}*sin(2\pi U_2)\\
\beta(U_1,U_2)=\sqrt{-2ln(U_1)}*cos(2\pi U_2)
\end{aligned}
\end{equation}
where $\alpha(U_1,U_2)$ and $\beta(U_1,U_2)$ are the random numbers following Gaussian distribution respectively.
\end{enumerate}

The amplitude of the random numbers {\toreferee $\alpha(U_1,U_2)$ and $\beta(U_1,U_2)$} depends on the uniform random number. Their phases equal to the product of $U_{1}$, $U_{2}$, and the constant 2$\pi$ .

{\toreferee
\subsection{Polarization Decision Algorithm}
The polarization decision algorithm method proposed by \citet{Bell:1968:ANR:363397.363547} is also a precise approach to obtain the two-dimensional Gaussian distribution. The polar algorithm is related to the Box-Muller transform but is superior to it.

Theoretically, we consider two independent and normally distributed random variables X and Y in Cartesian coordinates (i.e. $f_{X}(x)=\frac{1}{\sqrt{2\pi}\delta }e^{-\frac{x^2}{2\delta ^2}}$ and $f_{Y}(y)=\frac{1}{\sqrt{2\pi}\delta }e^{-\frac{y^2}{2\delta ^2}}$). Then the probability density function are $F=\int^{+\infty}_{-\infty} f_{X}(x)dx=\int^{+\infty}_{-\infty} f_{Y}(y)dy$. Thus the square of $F$ transformed to polar coordinate is:
\begin{equation}
F^{2}=\frac{1}{2\pi\delta^{2} }\int^{+\infty}_{-\infty}\int^{+\infty}_{-\infty}e^{-\frac{x^2+y^2}{2\delta ^2}}dxdy=\frac{1}{2\pi\delta^{2} }\int^{2\pi}_{0}\int^{+\infty}_{0}re^{-\frac{r^2}{2\delta ^2}}drd\theta
\end{equation}

Similarly to Box-Muller method, the transformation to polar coordinates makes $\theta$ is uniformly distributed from 0 to $2\pi$. The normalized distribution function of radial distance $r$ is:
\begin{equation}
P(r<a)=\int^{a}_{0}re^{-\frac{r^2}{2}}dr
\end{equation}

The uniform random number U is also used here. Since U is uniformly distributed in the interval (0,1), then the point $(cos(2\pi U), sin(2\pi U))$ is uniformly distributed on the unit circumference $x^{2}+y^{2} = 1$. A new point is generated by multiplying that point by radial distance $r$: $(rcos(2\pi U), rsin(2\pi U))$. Finally, by the inverse transform, one obtains two jointly distributed two variables which are independent standard normal random variables.

In practical, the polar method is achieved by the rejection approach. Assuming $y = f (x)$ is a function with finite integral, $C$ is a set of points $(x, y)$, and $Z$ is a superset of $C$. Then from set $Z$, random points $(x, y)$ are uniformly selected until point $(x, y)$ falls into the range of $C$. The selected point $(x, y)$ is returned as the random number\cite{Knuth:1997:ACP:270146}. The set C here is set as a unit cycle: $C=x^{2}+y^{2}<1$, where $-1 < x < 1$, $-1 < y < 1$.
Then the random points $(U_1, U_2)$ is selected until  $C=U_1^{2}+U_2^{2}<1$. Then a pair of normal random variables is obtained as\cite{thomas2007gaussian}:
\begin{equation}
\begin{aligned}
\alpha=U_1\sqrt{\frac{-2ln(U_1^{2}+U_2^{2})}{U_1^{2}+U_2^{2}}}\\
\beta=U_2\sqrt{\frac{-2ln(U_1^{2}+U_2^{2})}{U_1^{2}+U_2^{2}}}
\end{aligned}
\end{equation}

\subsection{Central Limit Algorithm}
The central limit algorithm is based on the central limit theorem which states that when a sufficiently large number of samples drawn from independent random variables (i.e., uniform distributions ), the arithmetic mean of their distributions will have a normal distribution. Thus, the central limit algorithm is an extremely efficient method in GRNs generation, since it simply samples sufficient amount of identical and independent uniform distributions.

More formally, assume there are n independent and identically distributed uniform numbers $U_{i}\sim U(0,1)$. Then we can approximate results of the sum of $U_{i}$ as: $S=\sum_{i=1}^{n}U_{i}$. The cumulative distribution function of $S$ can be approximated as:
\begin{equation}
F_{S}(s)=\Phi(\frac{s-n\mu}{\sqrt{n\sigma^2}})
\end{equation}
where $\Phi$ represents the cumulative distribution function of a Gaussian distribution. For a uniform random variable $U_{i}\sim U(0,1)$, the mean $\mu$ and variance $\sigma$ are given by $\frac{1}{2}$ and
$\frac{1}{12}$ respectively. Thus if we choose variable $z=\frac{s-\frac{n}{2}}{\frac{1}{12}\sqrt{n}} $ the distribution function of z is Gaussian distribution:
\begin{equation}
f_{Z}(z)=\frac{n}{\sqrt{2\pi}}e^{-\frac{x^2}{2}}
\end{equation}
After normalization, a standard Gaussian distribution is obtained. Central limit algorithm  can  be  used  to  transform uniform  random  numbers  to  Gaussian  with  a  very  low  hard-ware  cost.  However,  the error  in  tail  regions  of  is  inversely proportional  to  the  number  of  $U_{i}\sim U(0,1)$  to  be  added. This makes the GRNs produced by central limit algorithm is not highly accurate in tail region\cite{thomas2007gaussian}.
}
\section{Hardware Architecture of GRN and URN Algorithm}
\label{sec:4}
\subsection{Uniform Random Number Generator}
\begin{figure}[htbp]
\centering
\includegraphics[width=0.98\linewidth,height=0.30\linewidth]{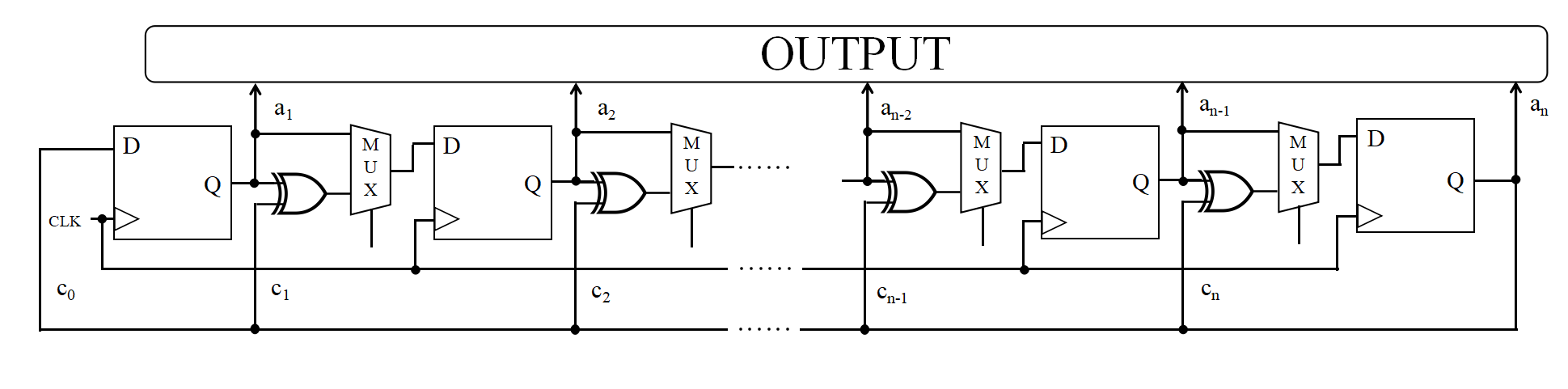}
\caption{General architecture of multi-return shift register generator. $c_{0}, c_{1}...c_{n-1},c_{n}$are the feedback coefficients and $a_{1}, a_{2}, a_{3}...a_{n}$are the output values.}
\label{fig:m}
\end{figure}
{\toreferee Since uniform random numbers (URNs) $U(0,1)$ is essential for all of the three algorithms introduced in Sec.\ref{sec:3}, an efficient and robust  URNs generator is indispensable in the whole system. Here we choose to use}  Multi-return Shift Register Generator (MSRG) \cite{Zhou2010Study}{\toreferee in our design.}

{\toreferee Multi-return Shift Register Generator is one type of } Linear Feedback Shift Register (LFSR)\cite{Xiao2009Multi} {\toreferee which} is one of the most effective and simple ways to get uniform random number. {\toreferee  The basic architecture of MSRG is  shown in Fig.\ref{fig:m}.  The MSRG is composed of a shift register and a feedback function, which can be represented as a polynomial of variable x referred to as the characteristic polynomial:}
\begin{equation}
f(x)=c_{n}x^n+c_{n-1}x^{n-1}+\cdots+c_{1}x+c_0
\end{equation}

 Where $c_{1}, c_{2}, c_{3}...c_{n}$ are the feedback coefficients. {\toreferee The feedback coefficients are selected by the multiplexer which is controlled by the control signal. When the multiplexer selects the output signal $a_{i}$ directly instead of the 'XOR' function, the corresponding coefficient $c_{i}$ would be regarded as 0 in characteristic polynomial.  The input bit is given from a linear function of the initial status and the next state of an MSRG is uniquely determined from the previous one by the feedback network. The initial value of the register is called seed and the sequence produced is completely determined by the initial status\cite{mioc2008complete}. Because the register has a finite number of possible statuses, after a period the sequence will be repeated. } The period of MSRG with order n is no more than $2^{n}-1$. Only if the feedback coefficients are properly chosen, the output sequence is the m sequence which has the longest period $2^{n}-1$ \cite{Zhang2014Algorithm}. In this paper, the primitive polynomial we choose is:

\begin{equation}
f(x)=x^{32}+x^{8}+x^{5}+x^{2}+1
\end{equation}

%In this work, the initial values of $a_{i}$ (the seed) {\toreferee is set as: 0000 0000 0000 0000 0000 0000 0000 0001 and 1111 1111 1111 1111 1111 1111 1111 1110.}
After getting m sequence with period of $(2^{32}-1)$, the uniform random number is obtained by dividing the m sequence by $(2^{n}-1)$ using divider.
{\toreferee The uniform random number generator which is the essential part of Gaussian random number generator has been obtained in Sec\ref{sec:4}. Then we show the hardware architecture design for implementing Box-Muller, polarization decision and central limit algorithm.}

\subsection{Hardware Architecture Deign for Box-Muller Algorithm}
\begin{figure}[htbp]
\centering
\includegraphics[width=0.99\linewidth,height=0.38\linewidth]{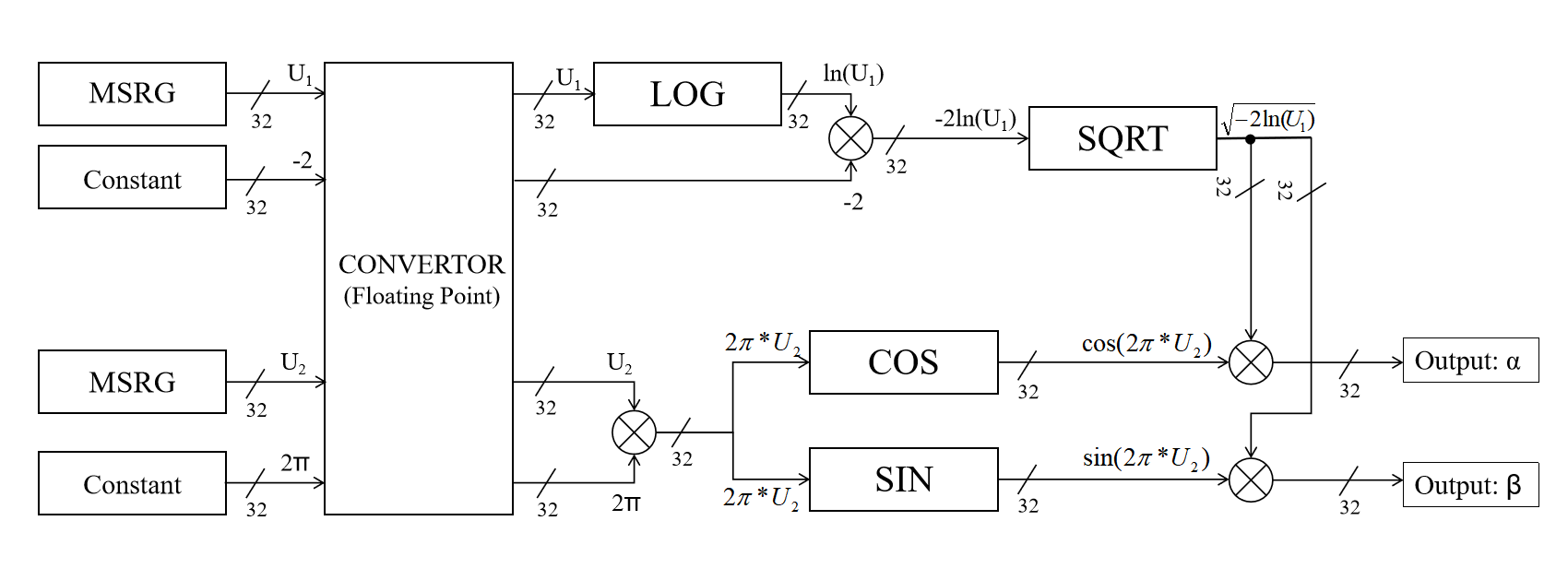}
\caption{\label{GRN1}The integral structural design diagram of Gaussian random generator,{\toreferee using Box-Muller algorithm. $U_{1}$ and $U_{2}$ are the uniform random numbers generated by MSRG}. $\alpha$ and $\beta$  are the output Gaussian random numbers. {\toreferee Sing-precision floating point numbers are used in the design.}}
\end{figure}
Fig.\ref{GRN1} shows the component structural design diagram of the Gaussian random generator, {\toreferee using the Box-Muller algorithm.
The two uniform random number $U_{1}$ and $U_{2}$ are generated by  MSRG module, which has been introduced in
Sec.\ref{sec:4}. All the random numbers $U_{1}$, $U_{2}$ and indispensable constants are converted into single-precision floating point numbers. The logarithm of  $U_{1}$ is achieved by 'LOG' module and the square root of $-2ln(U_{1})$ is calculated by module 'SQRT' (see Fig.\ref{GRN1}). The trigonometric functions $sin(2\pi U_{2})$ and $cos(2\pi U_{2})$ are achieved by modules 'COS' and 'SIN'. Four external multipliers are also used in the design. As a result, two sets of Gaussian random numbers are obtained, i.e., $\alpha$ and $\beta$.}

{\toreferee
\subsection{Hardware Architecture Deign for Polarization Decision Algorithm}
}{\toreferee Fig.\ref{GRN2} shows the component structural design diagram of the Gaussian random generator,using polarization decision algorithm.
The two uniform random number $U_{1}$ and $U_{2}$ are generated by  MSRG module, which has been introduced in  Sec.\ref{sec:4}. All the random numbers $U_{1}$, $U_{2}$ and indispensable constants are converted into single-precision floating point numbers. The logarithm of $U_{1}$ is achieved by 'LOG' module and the square root of $-2ln(U_{1})$ is calculated by module 'SQRT' (see Fig.\ref{GRN2}). The division function is achieved by modules 'DIV'. Five external multipliers and one adder are also used in the design. As a result, two sets of Gaussian random numbers are obtained\footnote{The output is always large than zero when $U_{1}, U_{2}\in(0,1)$. To get the bilateral Gaussian distribution, two addition sets of uniform random number $U_{3}, U_{4}\in(0,1)$ are indispensable. The key point here is to change the sign bit of $U_{3}, U_{4}$ to be negative after converted into floating point number, and then combine the results from  $U_{1}, U_{2},U_{3}, U_{4}$.}, i.e., $\alpha$ and $\beta$.
\begin{figure}[htbp]
\centering
\includegraphics[width=0.94\linewidth,height=0.45\linewidth]{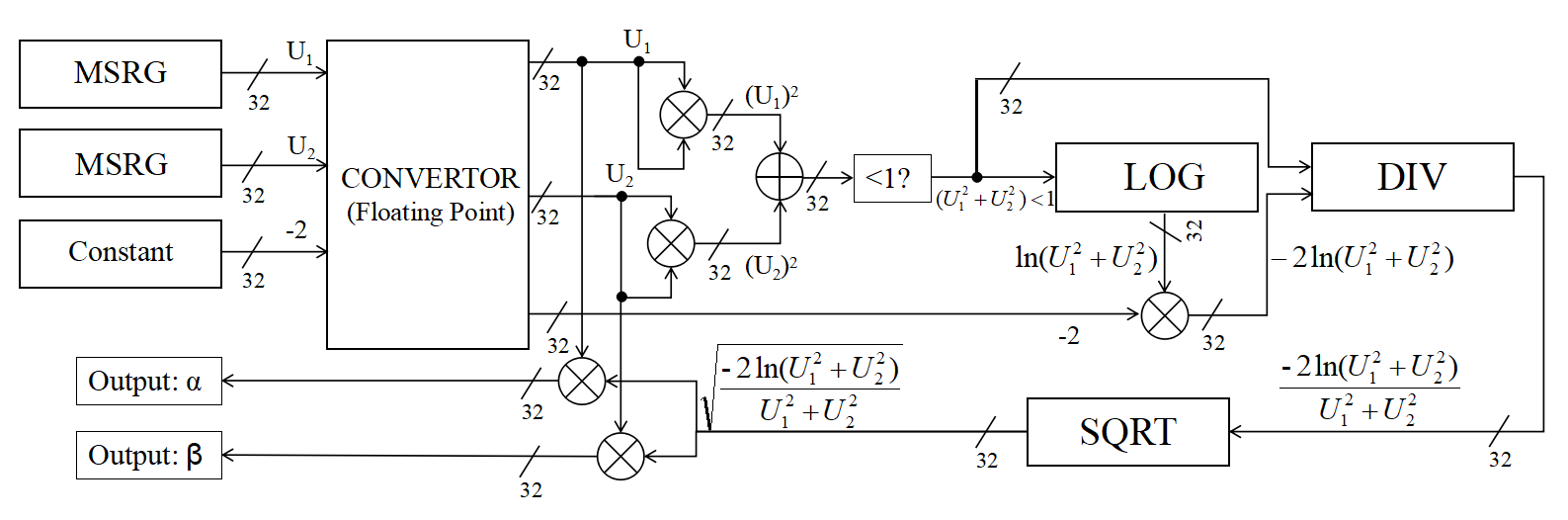}
\caption{\label{GRN2}{\toreferee The integral structural design diagram of Gaussian random generator, using polarization decision algorithm. $U_{1}$ and $U_{2}$ are the uniform random numbers generated by MSRG. $\alpha$ and $\beta$  are the output Gaussian random numbers. Sing-precision floating point numbers are used in the design.}}
\end{figure}

{\toreferee
\subsection{Hardware Architecture Deign for Central Limit Algorithm}
Fig.\ref{GRN3} shows the component structural design diagram of the Gaussian random generator,using central limit algorithm. The uniform random numbers $U_{1}, U_{2}...U_{n}$ are generated by  MSRG module, which has been introduced in Sec.\ref{sec:4}. All the random numbers $U_{1}, U_{2}...U_{n}$ and indispensable constants are converted into single-precision floating point numbers. The new variable  $S=\sum_{i=1}^{n}U_{i}$ is obtained by a adder and the square root is calculated by module 'SQRT' (see Fig.\ref{GRN3}). The division function is achieved by modules 'DIV'. Tow external multipliers and one adder are used in the design. As a result, one set of Gaussian random numbers is obtained, i.e., $\alpha$.
}

\begin{figure}[htbp]
\centering
\includegraphics[width=0.95\linewidth,height=0.48\linewidth]{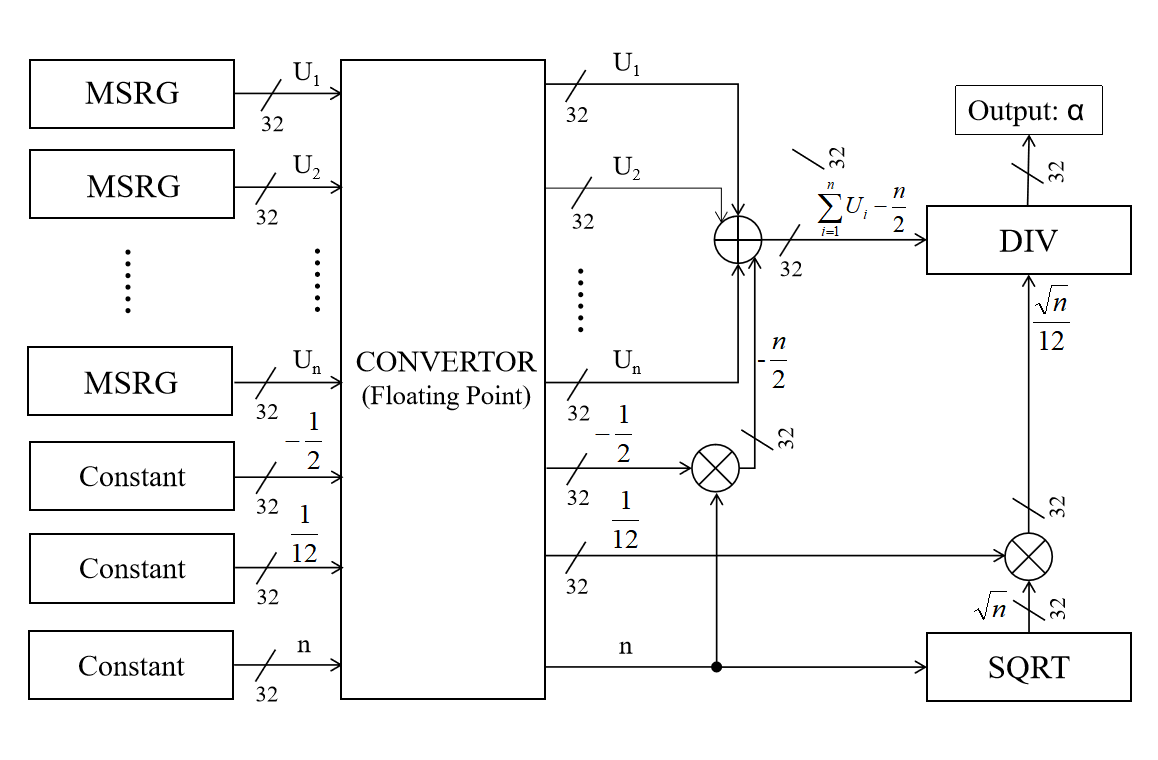}
\caption{\label{GRN3}{\toreferee The integral structural design diagram of Gaussian random generator, using central limit algorithm. $U_{1}, U_{2}...U_{n}$ are the uniform random numbers generated by MSRG. n is the number of uniform random numbers sets. $\alpha$ is the output Gaussian random numbers. Sing-precision floating point numbers are used in the design.}}
\end{figure}
}

\section{Hardware Realization of Algorithm Functions}
\label{sec:5}
\subsection{ Field-Programmable Gate Array}
{\toreferee The Field-Programmable Gate Array (FPGA), which integrates programmable logic blocks, soft-core or hardcore processors, has become more and more common as a core technology used to build electronic systems. In most FPGAs, logic blocks also include memory elements, which may be simple flip-flops or more complete blocks of memory. The FPGA configuration is generally specified using a hardware description language, like what we use in this work: Verilog HDL.

The main and the most significant difference between the micro-controller and the FPGA is that FPGA does not have a fixed hardware structure. On the contrary, FPGA is programmable according to user applications. However, processors have a fixed hardware structure, which means that all the transistors memory, peripheral structures, and the connections are constant. Which the processor predefine the operations (addition, multiplication, I/O control, etc.), and then users make the processor sequentially do these operations by using a software.

Hardware structure in the FPGA is not fixed but defined by the user. Although logic cells are fixed in FPGA, functions they perform and the interconnections between them are determined by the user. So operations that FPGA can do are not predefined. Users can have the processes done according to the written HDL code "in parallel" which means simultaneously. The ability of parallel processing is one of the most critical features that separate FPGA from the processor and make it superior in many areas.

FPGA is generally more useful for routine control of particular circuits. For example, using FPGA for simple functions such as check the quantum key signals from communication. This process can be quickly done with many conventional micro-controllers (PIC series, etc.). However, a solution from FPGA is more reasonable, if users want to achieve a high-efficient communication.

Because QCK processing requires processing large data in high speed and make these types of applications are very suitable for FPGA that is capable of parallel processing. Since the user can determine the hardware structure of FPGA, FPGA can be programmed to process more extensive data with few clock cycle. Whereas, it is not possible to achieve this performance by the processor. Because data flow is limited by processor bus (16-bit, 32 bit, etc.) and the processing speed. As a result, for applications that require more performance such as intensive data processing FPGA has come to the fore for routine control operations. Nevertheless, micro-controllers can be embedded into the FPGA since they are logic circuits in fact. Thus it possible to define and use processor and user-specific hardware functions on only one chip by using FPGA. This solution shows the possibility to control the hardware because of its high flexibility. Users can modify and update whole design (FPGA on the processor and other logic circuits) by only changing the code on FPGA, without any change on circuit board layout. In this way, users can add different functions, improve performance and make your design resistant to time without having to redesign the cards.}

According to the Gaussian random number generation algorithm described above, FPGA chip Altera Cyclone IV E EP4CE115F29I8L is chosen to achieve our design. 528 I/O ports, 114,480 logic elements, and 7155 logic array blocks are embedded in this chip. It is able to achieve 200 MHz maximum operating frequency.

\begin{figure}[htbp]
\centering

\includegraphics[width=0.7\linewidth,height=0.55\linewidth]{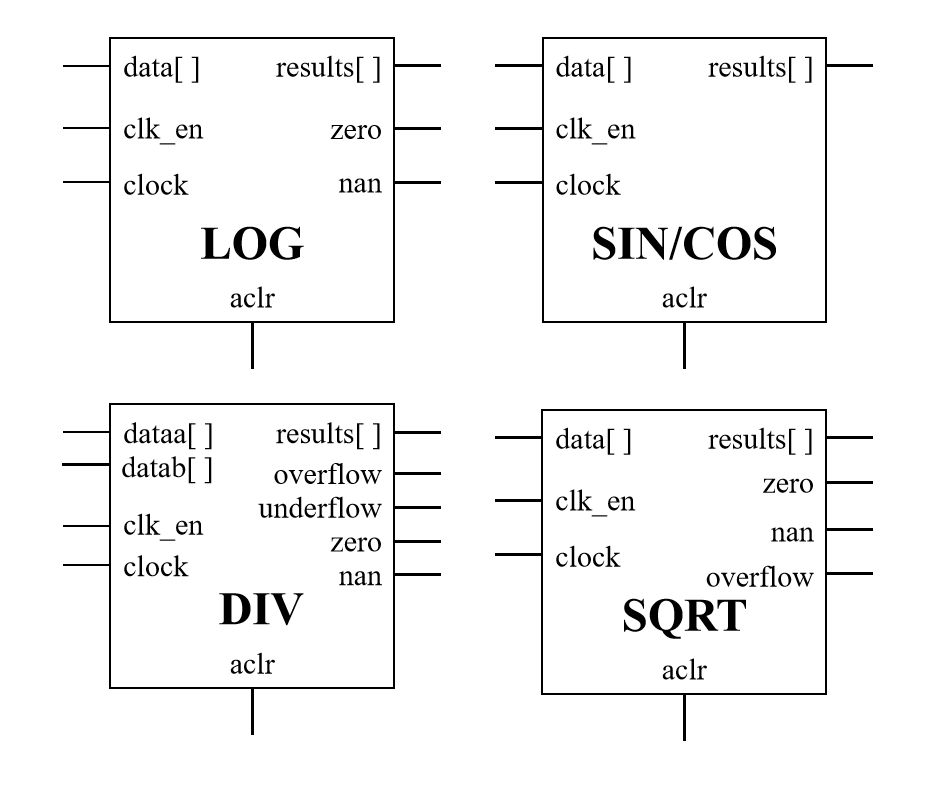}
\caption{\label{ip cores}The input and output signals of FPGA floating-point IP cores. LOG is ALTFP\_LOG IP core. SIN/COS is ALTFP\_SINCOS IP core. DIV is ALTFP\_DIV IP core. SQRT is ALTFP\_SQRT IP core. }
\end{figure}

{\toreferee
\subsection{FPGA Floating-Point IP Cores }
Intellectual property (IP) cores are standalone modules that can be used in any field programmable gate array and  source codes are ported across various FPGA platforms. These are developed using HDL languages like VHDL, Verilog and System Verilog. In this work we use soft IP cores to implement our design. Soft IP cores are completely flexible and do not depend on vendor technology. Hence, the IPs can be modified according to users' typical application and easily integrated with other modules.

\subsubsection{Logarithm IP Core}
The logarithm calculation is achieved by the ALTFP\_LOG IP core which can compute the natural logarithm of single-precision format numbers. Fig\ref{ip cores} shows the input and output signals of the ALTFP\_LOG IP core. The function of each port is defined as:
\begin{itemize}
\item \textbf{clock}: Clock input to the IP core;
\item \textbf{clk\_en}: Clock enable. When the \textbf{clk\_en} port is asserted high, a natural logarithm operation takes place.
\item \textbf{aclr}: Asynchronous clear. When the \textbf{aclr} port is asserted high, the function is asynchronously cleared.
\item \textbf{data[ ]}: Floating-point input data.
\item \textbf{result[ ]}: The natural logarithm of the value on input data.
\item \textbf{zero}: Zero exception output. This occurs when the actual input value is 1.
\item \textbf{nan}: NaN exception output. This occurs when the input is a negative number or NaN.
\end{itemize}

\subsubsection{Trigonometric IP Core}
The trigonometric calculation is achieved by the ALTFP\_SINCOS IP core which can perform trigonometric sine and cosine functions single-precision format numbers. Fig\ref{ip cores} shows the input and output signals of the ALTFP\_SINCOS IP core. The function of each port is defined as:

\begin{itemize}
\item \textbf{clock}: Clock input to the mega-function.;
\item \textbf{clk\_en}: Clock enable. When the \textbf{clk\_en} port is asserted high, sine or cosine operation takes place.
\item \textbf{aclr}: Asynchronous clear. When the \textbf{aclr} port is asserted high, the function is asynchronously cleared.
\item \textbf{data[ ]}: Floating-point input data.
\item \textbf{result[ ]}: The trigonometric of the \textbf{data[]} input port in floating-point format. The widths of the \textbf{result[]} output port and \textbf{data[]} input port are the same.
\end{itemize}

\subsubsection{Division IP Core}
The division is achieved by the ALTFP\_DIV IP core which performs the floating-point division operation. Fig\ref{ip cores} shows the input and output signals of the ALTFP\_DIV IP core. The function of each port is defined as:

\begin{itemize}
\item \textbf{clock}: Clock input to the IP core;
\item \textbf{clk\_en}: Clock enable to the floating-point divider. This port enables division.
\item \textbf{aclr}: Asynchronous clear. When the \textbf{aclr} port is asserted high, the function is asynchronously cleared.
\item \textbf{dataa[ ]}: Numerator data input.
\item \textbf{datab[ ]}: Denominator data input.
\item \textbf{result[ ]}: Divider output port. The division result.
\item \textbf{overflow}: Overflow port for the divider. Asserted when the result of the division exceeds or reaches infinity.
\item \textbf{underflow}: Underflow port for the divider. Asserted when the result of the division is zero even though neither of the inputs to the divider is zero, or when the result is a denormalized number.
\item \textbf{zero}: Zero port for the divider. Asserted when the value of \textbf{result[]} is zero.
\item \textbf{nan}: NaN port. Asserted when an invalid division occurs, such as infinity dividing infinity or zero dividing zero.
\end{itemize}

\subsubsection{Square Root Calculation IP Core}
The square root calculation is achieved by the ALTFP\_SQRT IP core. This IP core performs a square root calculation based on the input provided. Fig\ref{ip cores} shows the input and output signals of the ALTFP\_SQRT IP core. The function of each port is defined as:

\begin{itemize}
\item \textbf{clock}: Clock input to the IP core;
\item \textbf{clk\_en}: Clock enable that allows square root operations when the port is asserted high.
\item \textbf{aclr}: Asynchronous clear. When the \textbf{aclr} port is asserted high, the function is asynchronously cleared.
\item \textbf{data[ ]}: Floating-point input data.
\item \textbf{result[ ]}: Square root output port for the floating-point result.
\item \textbf{zero}: Zero port. Asserted when the value of the \textbf{result[]} port is 0.
\item \textbf{nan}: NaN port. Asserted when an invalid square root occurs, such as negative numbers or NaN inputs.
\item \textbf{overflow}: Overflow port. Asserted when the result of the square root exceeds or reaches infinity.
\end{itemize}

\section{The Result of Simulation and Testing}
\label{sec:6}

\subsection{FPGA Resource Usage Summary}
\begin{table}[htbp]
\centering
\begin{tabular}{|c|c|c|c|c|}
\hline
Algorithm & <= 2 input functions & 3 input function & 4 input function & I/O pins \\ \hline
Box-Muller & 1990 & 6768 & 4335 & 131 \\ \hline
Polarization Decision & 1579 & 2859 & 1469 & 131 \\ \hline
Central Limit   & 1453 & 2873 & 2889 & 420 \\ \hline
 \end{tabular}
\caption{\label{Tab:LUT}Logic element usage by number of LUT input and I/O pins. The table shows the LUT resource usage for achieving Box-Muller algorithm, polarization decision algorithm, and central limit algorithm respectively.}
\end{table}
Based on the design of three algorithms above, we achieve the function by using FPGA. Here we show the resource utilization of LUT, logic elements, and fan-out analysis for each design. Please note, we choose 12 sets of uniform random number as inputs to the central limit algorithm.

Tab.\ref{Tab:LUT} presents the logic element usage by the number of LUT input and I/O pins. The design of the Box-Muller algorithm requires more LUT resources than the others, because of its complex mathematical operations. Moreover, Polarization decision algorithm not only uses less LUT resource, in particular, the four input function but also less I/O pins.

Tab.\ref{Tab:mode} shows the logic element usage by normal mode and arithmetic mode. The design of the central limit algorithm requires fewer arithmetic resources than the others because it does not involves with logarithm and trigonometric. Moreover, the Box-Muller algorithm still uses much more resources.

\begin{table}[htbp]
\centering
\begin{tabular}{|c|c|c|}
\hline
Algorithm & Logic Elements by Normal Mode & Logic Elements by Arithmetic Mode  \\ \hline
Box-Muller & 7803 & 5209  \\ \hline
Polarization Decision & 3602 & 2305  \\ \hline
Central Limit   & 6142 & 1073  \\ \hline
 \end{tabular}
\caption{\label{Tab:mode}Logic element usage by mode. The table shows the logic elements resource usage for achieving Box-Muller algorithm, polarization decision algorithm, and central limit algorithm respectively.}
\end{table}

Tab.\ref{Tab:fanout} gives the results of fan-out rest. The maximum fan-out and total fan-out of Box-muller algorithm is much large than the others. Polarization decision and central limit algorithm gives similar results of maximum fan-out and total fan-out values. However, the central limit algorithm shows a large average fan-out than polarization decision.
\begin{table}[htbp]
\centering
\begin{tabular}{|c|c|c|c|}
\hline
Algorithm & Maximum Fan-Out & Total Fan-Out & Average Fan-Out\\ \hline
Box-Muller & 8790 & 62386 & 2.81 \\ \hline
Polarization Decision & 5262 & 30332 & 2.63  \\ \hline
Central Limit  & 4534 & 35411 & 2.81 \\ \hline
 \end{tabular}
\caption{\label{Tab:fanout}The table shows the results of fan-out test using Box-Muller algorithm, polarization decision algorithm, and central limit algorithm respectively.}
\end{table}
\begin{figure*}
\centering
\subfigure[Box-Muller: $\alpha$ set
]
{\includegraphics[width=.49\linewidth,height=0.35\linewidth]{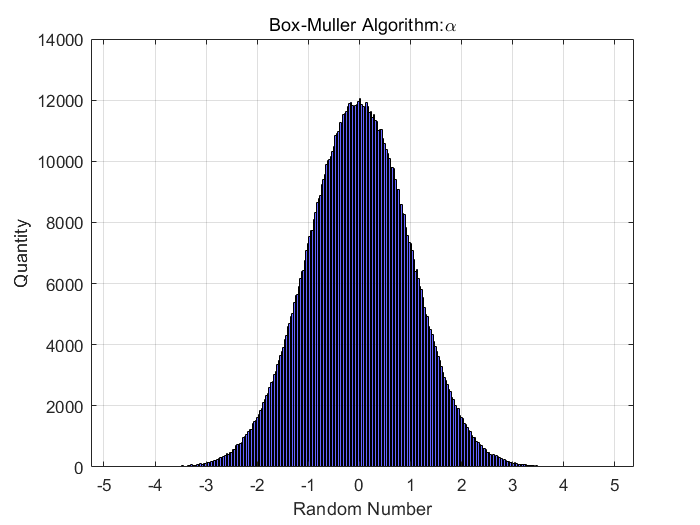}}
\centering
\subfigure[Box-Muller: $\beta$ set
]
{\includegraphics[width=.49\linewidth,height=0.35\linewidth]{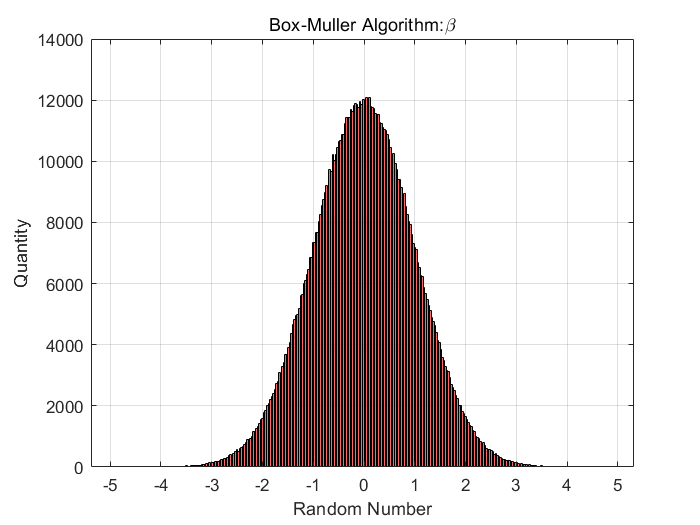}}
\centering
\subfigure[Polarization decision: $\alpha$ set
]
{\includegraphics[width=.49\linewidth,height=0.35\linewidth]{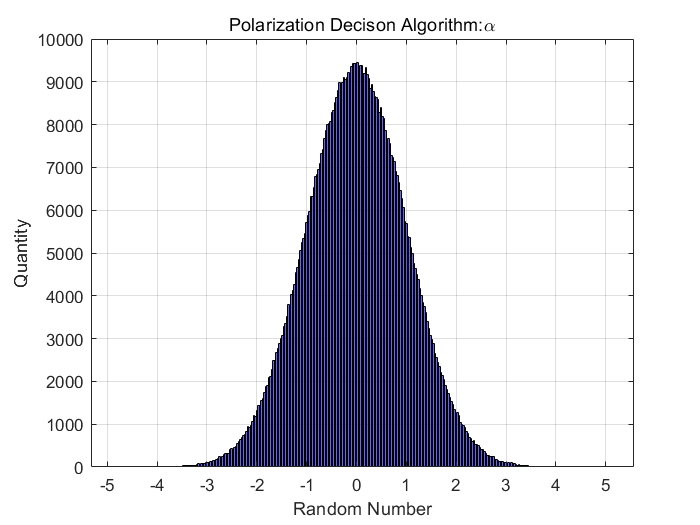}}
\centering
\subfigure[Polarization decision: $\beta$ set
]
{\includegraphics[width=.49\linewidth,height=0.35\linewidth]{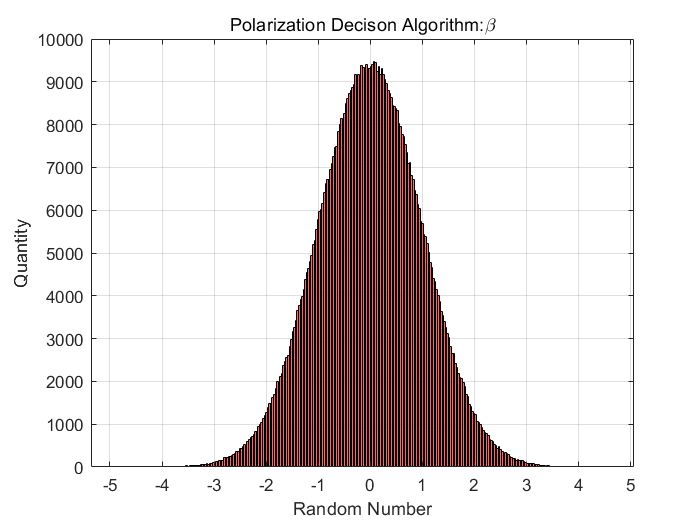}}
\centering
\subfigure[Central Limit
]
{\includegraphics[width=.49\linewidth,height=0.35\linewidth]{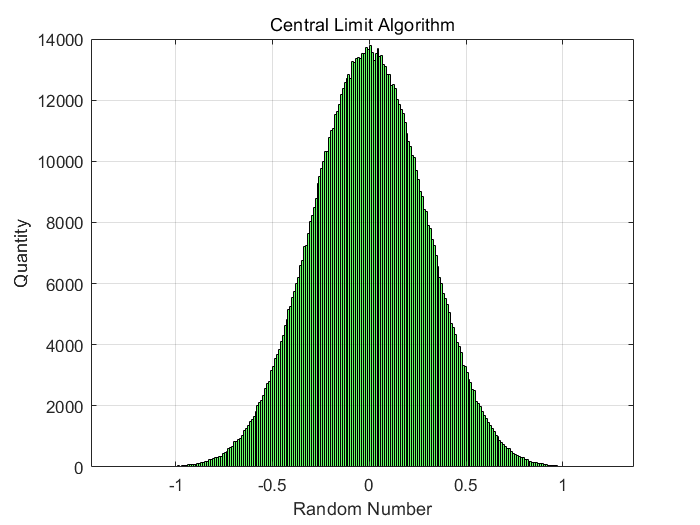}}
\centering
\subfigure[MATLAB
]
{\includegraphics[width=.49\linewidth,height=0.35\linewidth]{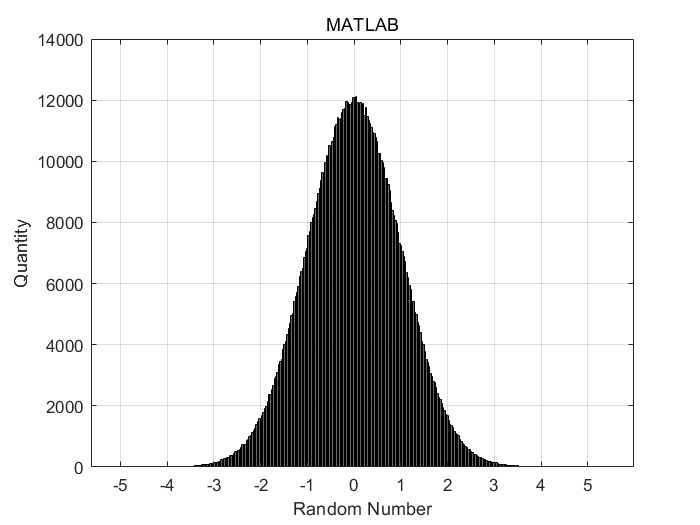}}
\caption{\label{fig:hist}The histogram of Gaussian random number generated by Box-Muller, polarization decision, and central limit algorithm using FPGA, as well as the random number generated by MATLBE software function 'randn'.}
\end{figure*}
\subsection{Statistical Analysis}
To verify our design of the three algorithms, we import the random number generated to MATLAB for statistical analysis and compare the results with the random number generated by the function 'randn' \cite{nadler2006design} MATLAB software. The number of sampled random number is 1,000,000 for all cases.

Fig.\ref{fig:hist} shows the histogram of random numbers generated by Box-Muller, polarization decision, and central limit algorithm. It is evident that all sets of random numbers follow the Gaussian profile. Box-Muller and polarization decision method are generating two sets of the random number simultaneously, i.e., $\alpha$ set and $\beta$ set. The quantity of random number generated by the polarization decision algorithm is less than the others since the two uniform random number $U_{1}$ and $U_{2}$ are rejected if $U_{1}^{2}+U_{2}^{2}>1$.

A more robust way to estimate the accuracy of Gaussian random numbers is the null hypothesis, which is used to determine what outcomes of a study would lead to a rejection of the null hypothesis for a pre-specified level of significance\cite{moore2012introduction}. We show the results of three null hypothesis in the following, i.e., Chi-Square Goodness of fit test, Anderson-Darling test, and Kolmogorov-Smirnov test.

\subsection{The Null Hypothesis Test}
The Chi-squared test is a statistical test, which is a null hypothesis stating that the frequency distribution of specific events observed in a sample is consistent with a particular theoretical distribution. The theoretical distribution here is the chi-squared distribution. This test suitable for unpaired data from large samples \cite{lani2011chi}. The significance level is set to be 5\% here.

The results are presented in Tab.\ref{Tab:cs_test}. The central limit method is rejected by the null hypothesis. It indicates the approximation the Gaussian is particularly
poor, especially in the tails\cite{thomas2007gaussian}. To improve the accuracy of the central limit method, one has to increase the number of the uniform random number used for approximation(See Sec.\ref{sec:4}). However, large numbers of uniform random numbers, also, constitutes a computational challenge. Thus, the central limit theorem is not ideally used in contemporary GRNs. The Box-Muller method and polarization method both pass the Chi-squared test. Also, $\beta$ set of Box-Muller shows a smaller P value than the one generated by MATLAB. It indicates the quality of GRNs generated by the Box-Muller algorithm in FPGA is higher than MATLAB software.

\begin{table}[htbp]
\centering
\begin{tabular}{|c|c|c|c|}
\hline
Algorithm & Null Hypothesis & Calculated Probability (P Value) & Test Statistic\\ \hline
Box-Muller & Non-rejected & $\alpha$: 0.6636, $\beta$:0.1238  & $\alpha$: 4.9290, $\beta$:11.3567\\ \hline
Polarization Decision & Non-rejected & $\alpha$: 0.7312, $\beta$:0.8725 & $\alpha$: 4.4130, $\beta$:3.1322 \\ \hline
Central Limit  & Rejected & NaN & NaN  \\ \hline
MATLATB & Non-rejected & 0.1938 & 9.9094 \\ \hline
 \end{tabular}
\caption{\label{Tab:cs_test}The table shows the results of the Chi-Square goodness of fit test for the Gaussian random number generated by Box-Muller algorithm, polarization decision algorithm, central limit algorithm, and MATLAB software respectively.}
\end{table}
\begin{table}[htbp]
\centering
\begin{tabular}{|c|c|c|c|}
\hline
Algorithm & Null Hypothesis & Calculated Probability (P Value) &  Test Statistic\\ \hline
Box-Muller & Non-rejected & $\alpha$: 0.7663, $\beta$:0.0704  & $\alpha$: 0.2449, $\beta$:0.6922\\ \hline
Polarization Decision & Non-rejected & $\alpha$: 0.0862, $\beta$:0.9399 & $\alpha$: 0.6569, $\beta$:0.1736 \\ \hline
Central Limit  & Rejected & NaN & NaN  \\ \hline
MATLATB & Non-rejected & 0.3044 & 0.4339 \\ \hline
 \end{tabular}
\caption{\label{Tab:ad_test}The table shows the results of the Anderson-Darling test for the Gaussian random number generated by Box-Muller algorithm, polarization decision algorithm, central limit algorithm, and MATLAB software respectively.}
\end{table}

\begin{table}[htbp]
\centering
\begin{tabular}{|c|c|c|c|}
\hline
Algorithm & Null Hypothesis & Calculated Probability (P Value) &  Test Statistic\\ \hline
Box-Muller & Non-rejected & $\alpha$: 0.5481, $\beta$:0.3308  & $\alpha$: 0.0008, $\beta$: 0.0010 \\ \hline
Polarization Decision & Non-rejected & $\alpha$: 0.1778, $\beta$:0.4333 & $\alpha$: 0.0012, $\beta$:0.0010 \\ \hline
Central Limit  & Rejected & NaN & NaN  \\ \hline
MATLATB & Rejected & NaN & NaN \\ \hline
 \end{tabular}
\caption{\label{Tab:ks_test}The table shows the results of the one-sample Kolmogorov-Smirnov test for the Gaussian random number generated by Box-Muller algorithm, polarization decision algorithm, central limit algorithm, and MATLAB software respectively.}
\end{table}
Furthermore, the Anderson-Darling test and the Kolmogorov-Smirnov test are also applied here for estimation. Anderson-Darling test is a statistical test to confirm whether a given sample of data is drawn from a given probability distribution\cite{scholz1987k}, while the Kolmogorov-Smirnov test is a nonparametric test of the equality of continuous probability distributions that can be used to compare a sample with a reference probability distribution\cite{massey1951kolmogorov}. When applied to test whether a normal distribution adequately describes a set of data, they are the most powerful statistical tools for detecting most departures from normality.

The results are shown in Tab.\ref{Tab:ad_test} and Tab.\ref{Tab:ks_test}. The central limit method is still rejected by these two test. More important, the GRNs generated by MATLAB function 'randn' does not pass the Kolmogorov-Smirnov test, while the Box-Muller method and polarization method both pass the Kolmogorov-Smirnov test. These results confirm that our design for GRNs using Box-Muller and polarization decision algorithm is a robust way to generate high-quality GRNs.

\section{Discussion}
\label{sec:7}
Two groups of the Gaussian random number generated through Box-Muller and polarization decision algorithm have been examined to be well-distributed with high quality. To apply our GRNs design to a QKD system, one addition electrical modulator is required. Through the modulator, the phase and magnitude signal which follow Gaussian profile is modulated. The signal is regarded as pseudo quantum states, which usually generated by expensive optical devices.

In a QKD system, high speed and efficient communication between two parties are required. The ability of parallel processing of FPGA shows an advantage in this case. Since the user can determine the hardware structure of FPGA, FPGA can be programmed to process more extensive data with few clock cycle. The high-speed communication is achieved. To guarantee the security of communication, usually, a truly
random signal is required. However, the GRNs generated FPGA can be regarded as true GRNs when its period is long enough. Because of the extensive and flexible resources of FPGA, the period of GRNs can be extremely long. Of course, the extended period required a better FPGA type which requires more investment.

}

\section{Conclusion}
\label{sec:8}
{\toreferee
Quantum Key Distribution is the process of using quantum communication to establish a shared key between two parties. The absolute security and effective communication of quantum communication system can be guaranteed by a good Gaussian random number generator with high speed and a long random period. In this paper, we propose a possible scheme whose results are proved to be satisfactory and all of these works are the foundation of subsequent work. We conclude that:
\begin{enumerate}
\item{The unfixed hardware structure of FPGA provides users the parallel processing solution and makes FPGA superior in many areas than the microprocessor. FPGA is an ideal solution for QCK processing which requires processing large data in high speed.}
\item{Among three conventional GRNs algorithms, the Gaussian random number generated through polarization decision algorithm shows higher quality than others.}
\item{FPGA floating-point IP cores can be easily modified and integrated with other modules. They are appropriate choices to achieve the complex mathematical operation in hardware.}
\end{enumerate}

}

\section*{Acknowledgement}
AL acknowledges the support of National Undergraduate Innovation Program 2016. We are grateful to Prof. Guochun Wan and Prof. Meisong Tong,
in the Department of Electronics and Information Engineering, Tongji University. We are extremely thankful and indebted
to them for sharing expertise, and sincere and valuable guidance and encouragement extended to us.
%\printendnotes

% Submissions are not required to reflect the precise reference formatting of the journal (use of italics, bold etc.), however it is important that all key elements of each reference are included.
%\begin{biography}[example-image-1x1]{A.~One}
%Please check with the journal's author guidelines whether author biographies are required. They are usually only included for review-type articles, and typically require photos and brief biographies (up to 75 words) for each author.
%\bigskip
%\bigskip

%\bibliography{sample}

\begin{thebibliography}{40}
\providecommand{\natexlab}[1]{#1}
\providecommand{\url}[1]{\texttt{#1}}
\providecommand{\urlprefix}{}










































\bibitem[{Argyris et~al.(2012)Argyris, Apostolos and Pikasis, Evangelos and Deligiannidis, Stavros and Syvridis, Dimitris}]{Argyris2012Sub} Argyris A, Pikasis E, Deligiannidis S, Syvridis D.\newblock Sub-Tb/s Physical Random Bit Generators Based on Direct Detection of Amplified Spontaneous Emission Signals. \newblock Journal of Lightwave Technology 2012;30(9):1329--1334.
\bibitem[{Bell(1968)Bell, James R.}]{Bell:1968:ANR:363397.363547} Bell JR. \newblock Algorithm 334: Normal Random Deviates. \newblock Commun ACM 1968 Jul;11(7):498--. \newblock \urlprefix\url{http://doi.acm.org/10.1145/363397.363547}.
\bibitem[{Bennett and Shor(1998)Bennett, Charles H and Shor, Peter W}]{bennett1998quantum} Bennett CH, Shor PW. \newblock Quantum information theory. \newblock to appear 1998;.
\bibitem[{Bouchiat et~al.(1998)Bouchiat, Vincent and Vion, D and Joyez, Ph and Esteve, D and Devoret, MH}]{bouchiat1998quantum} Bouchiat V, Vion D, Joyez P, Esteve D, Devoret M. \newblock Quantum coherence with a single Cooper pair. \newblock Physica Scripta 1998;1998(T76):165.
\bibitem[{Box and Muller(1958)Box, G. E. P. and Muller, Mervin E.}]{box1958} Box GEP, Muller ME. \newblock A Note on the Generation of Random Normal Deviates. \newblock Ann Math Statist 1958 06;29(2):610--611. \newblock \urlprefix\url{https://doi.org/10.1214/aoms/1177706645}.
\bibitem[{Braunstein and Van~Loock(2005)Braunstein, Samuel L and Van Loock, Peter}]{braunstein2005quantum} Braunstein SL, Van~Loock P. \newblock Quantum information with continuous variables. \newblock Reviews of Modern Physics 2005;77(2):513.
\bibitem[{Busch et~al.(2007)Busch, Paul and Heinonen, Teiko and Lahti, Pekka}]{busch2007heisenberg} Busch P, Heinonen T, Lahti P. \newblock Heisenberg's uncertainty principle. \newblock Physics Reports 2007;452(6):155--176.
\bibitem[{Cerf et~al.(2001)Cerf, Nicolas J and Levy, Marc and Van Assche, Gilles}]{cerf2001quantum} Cerf NJ, Levy M, Van~Assche G. \newblock Quantum distribution of Gaussian keys using squeezed states. \newblock Physical Review A 2001;63(5):052311.
\bibitem[{Chapuran et~al.(2009)Chapuran, TE and Toliver, P and Peters, NA and Jackel, J and Goodman, MS and Runser, RJ and McNown, SR and Dallmann, N and Hughes, RJ and McCabe, KP and others}]{chapuran2009optical} Chapuran T, Toliver P, Peters N, Jackel J, Goodman M, Runser R, et~al. \newblock Optical networking for quantum key distribution and quantum communications. \newblock New Journal of Physics 2009;11(10):105001.
\bibitem[{Deutsch(1985)Deutsch, David}]{deutsch1985quantum} Deutsch D. \newblock Quantum theory, the Church--Turing principle and the universal quantum computer. \newblock Proc R Soc Lond A 1985;400(1818):97--117.
\bibitem[{Gabriel et~al.(2010)Gabriel, Christian and Wittmann, Christoffer and Sych, Denis and Dong, Ruifang and Mauerer, Wolfgang and Andersen, Ulrik L. and Marquardt, Christoph and Leuchs, Gerd}]{Gabriel2010A} Gabriel C, Wittmann C, Sych D, Dong R, Mauerer W, Andersen UL, et~al. \newblock A generator for unique quantum random numbers based on vacuum states. \newblock Nature Photonics 2010;4(10):711--715.
\bibitem[{Grosshans et~al.(2003)Grosshans, Fr??d??ric and Assche, Gilles Van and Wenger, J??r?Me and Brouri, Rosa and Cerf, Nicolas J. and Grangier, Philippe}]{grosshans2003quantum} Grosshans F, Assche GV, Wenger J, Brouri R, Cerf NJ, Grangier P. \newblock Quantum key distribution using gaussian-modulated coherent states. \newblock Nature 2003;421(6920):238.
\bibitem[{Gruska(1999)Gruska, Jozef}]{gruska1999quantum} Gruska J. \newblock Quantum computing, vol. 2005. \newblock McGraw-Hill London; 1999.
\bibitem[{Herrerocollantes and Garciaescartin(2017)Herrerocollantes, Miguel and Garciaescartin, Juan Carlos}]{Herrerocollantes2017Quantum} Herrerocollantes M, Garciaescartin JC. \newblock Quantum Random Number Generators. \newblock Review of Modern Physics 2017;(1).
\bibitem[Hu et al.(2018)]{2018MNRAS.480.1333H} Hu, Y., Yuen, K.~H., Lazarian, A.\ 2018.\ Improving the accuracy of magnetic field tracing by velocity gradients: principal component analysis.\ Monthly Notices of the Royal Astronomical Society 480, 1333.
\bibitem[{Jofre et~al.(2011)Jofre, M and Curty, M and Steinlechner, F and Anzolin, G and Torres, J. P. and Mitchell, M. W. and Pruneri, V}]{Jofre2011True} Jofre M, Curty M, Steinlechner F, Anzolin G, Torres JP, Mitchell MW, et~al. \newblock True random numbers from amplified quantum vacuum. \newblock Optics Express 2011;19(21):20665.
\bibitem[{Kestur et~al.(2010)Kestur, Srinidhi and Davis, John D and Williams, Oliver}]{kestur2010blas} Kestur S, Davis JD, Williams O. \newblock Blas comparison on fpga, cpu and gpu. \newblock In: VLSI (ISVLSI), 2010 IEEE computer society annual symposium on IEEE; 2010. p. 288--293.
\bibitem[{Knuth(1997)Knuth, Donald E.}]{Knuth:1997:ACP:270146} Knuth DE. \newblock The Art of Computer Programming, Volume 2 (3rd Ed.): Seminumerical Algorithms. \newblock Boston, MA, USA: Addison-Wesley Longman Publishing Co., Inc.; 1997.
\bibitem[{Lani(2011)Lani, DJ}]{lani2011chi} Lani D, Chi-Square goodness of fit test. \newblock Retrieved january; 2011.
\bibitem[{Laudenbach et~al.(2017)Laudenbach, Fabian and Pacher, Christoph and Fung, Chi-Hang Fred and Poppe, Andreas and Peev, Momtchil and Schrenk, Bernhard and Hentschel, Michael and Walther, Philip and H{\"u}bel, Hannes}]{laudenbach2017continuous} Laudenbach F, Pacher C, Fung CHF, Poppe A, Peev M, Schrenk B, et~al. \newblock Continuous-Variable Quantum Key Distribution with Gaussian Modulation-The Theory of Practical Implementations. \newblock arXiv preprint arXiv:170309278 2017;.
\bibitem[{Lo et~al.(2014)Lo, Hoi-Kwong and Curty, Marcos and Tamaki, Kiyoshi}]{lo2014secure} Lo HK, Curty M, Tamaki K. \newblock Secure quantum key distribution. \newblock Nature Photonics 2014;8(8):595.
\bibitem[{Madsen et~al.(2012)Madsen, Lars S and Usenko, Vladyslav C and Lassen, Mikael and Filip, Radim and Andersen, Ulrik L}]{madsen2012continuous} Madsen LS, Usenko VC, Lassen M, Filip R, Andersen UL. \newblock Continuous variable quantum key distribution with modulated entangled states. \newblock Nature communications 2012;3:1083.
\bibitem[{Malik and Hemani(2016)Malik, Jamshaid Sarwar and Hemani, Ahmed}]{Malik:2016:GRN:2988524.2980052} Malik JS, Hemani A. \newblock Gaussian Random Number Generation: A Survey on Hardware Architectures. \newblock ACM Comput Surv 2016 Nov;49(3):53:1--53:37. \newblock \urlprefix\url{http://doi.acm.org/10.1145/2980052}.
\bibitem[{Massey~Jr(1951)Massey Jr, Frank J}]{massey1951kolmogorov} Massey~Jr FJ. \newblock The Kolmogorov-Smirnov test for goodness of fit. \newblock Journal of the American statistical Association 1951;46(253):68--78.
\bibitem[{Milonni(2013)Milonni, Peter W}]{milonni2013quantum} Milonni PW. \newblock The quantum vacuum: an introduction to quantum electrodynamics. \newblock Academic press; 2013.
\bibitem[{Mioc(????)Mioc, Mirella Amelia}]{mioc2008complete} Mioc MA. \newblock A complete analyze of using Shift Registers in Cryptosystems for Grade 4, 8 and 16 Irreducible Polynomials???.
\bibitem[{Moore et~al.(2012)Moore, David S and Craig, Bruce A and McCabe, George P}]{moore2012introduction} Moore DS, Craig BA, McCabe GP. \newblock Introduction to the Practice of Statistics. \newblock WH Freeman; 2012.
\bibitem[{Nadler(2006)Nadler, Boaz}]{nadler2006design} Nadler B. \newblock Design flaws in the implementation of the Ziggurat and Monty Python methods (and some remarks on Matlab randn). \newblock arXiv preprint math/0603058 2006;.
\bibitem[{Scarani et~al.(2009)Scarani, Valerio and Bechmann-Pasquinucci, Helle and Cerf, Nicolas J and Du{\v{s}}ek, Miloslav and L{\"u}tkenhaus, Norbert and Peev, Momtchil}]{scarani2009security} Scarani V, Bechmann-Pasquinucci H, Cerf NJ, Du{\v{s}}ek M, L{\"u}tkenhaus N, Peev M. \newblock The security of practical quantum key distribution. \newblock Reviews of modern physics 2009;81(3):1301.
\bibitem[{Scholz and Stephens(1987)Scholz, Fritz W and Stephens, Michael A}]{scholz1987k} Scholz FW, Stephens MA. \newblock K-sample Anderson--Darling tests. \newblock Journal of the American Statistical Association 1987;82(399):918--924.
\bibitem[{Shor and Preskill(2000)Shor, Peter W and Preskill, John}]{shor2000simple} Shor PW, Preskill J. \newblock Simple proof of security of the BB84 quantum key distribution protocol. \newblock Physical review letters 2000;85(2):441.
\bibitem[{Sidhu and Prasanna(2001)Sidhu, Reetinder and Prasanna, Viktor K}]{sidhu2001fast} Sidhu R, Prasanna VK. \newblock Fast regular expression matching using FPGAs. \newblock In: Field-Programmable Custom Computing Machines, 2001. FCCM'01. The 9th Annual IEEE Symposium on IEEE; 2001. p. 227--238.
\bibitem[{Thomas et~al.(2007)Thomas, David B and Luk, Wayne and Leong, Philip HW and Villasenor, John D}]{thomas2007gaussian} Thomas DB, Luk W, Leong PH, Villasenor JD. \newblock Gaussian random number generators. \newblock ACM Computing Surveys (CSUR) 2007;39(4):11.
\bibitem[{Thomas et~al.(2009)Thomas, David Barrie and Howes, Lee and Luk, Wayne}]{Thomas2009A} Thomas DB, Howes L, Luk W. \newblock A comparison of CPUs, GPUs, FPGAs, and massively parallel processor arrays for random number generation. \newblock In: Acm/sigda International Symposium on Field Programmable Gate Arrays; 2009. p. 63--72.
\bibitem[{Underwood(2004)Underwood, Keith}]{underwood2004fpgas} Underwood K. \newblock FPGAs vs. CPUs: trends in peak floating-point performance. \newblock In: Proceedings of the 2004 ACM/SIGDA 12th international symposium on Field programmable gate arrays ACM; 2004. p. 171--180.
\bibitem[{Wang(2006)Wang, L.Z.}]{Wang2006} Wang LZ. \newblock The Generation of Gaussian random number with FPGA and its Application on Quantum Crytography. \newblock In: The Generation of Gaussian random number with FPGA and its Application on Quantum Crytography; 2006. .
\bibitem[{Williams et~al.(2010)Williams, Caitlin R. S. and Salevan, Julia C. and Li, Xiaowen and Roy, Rajarshi and Murphy, Thomas E.}]{Williams2010Fast} Williams CRS, Salevan JC, Li X, Roy R, Murphy TE. \newblock Fast physical random number generator using amplified spontaneous emission. \newblock Optics Express 2010;18(23):23584.
\bibitem[{Xiao-Chen and Zhang(2009)Xiao-Chen, G. U. and Zhang, Min Xuan}]{Xiao2009Multi} Xiao-Chen GU, Zhang MX. \newblock Multi-output Fibonacci Type LFSR Based Uniform Random Number Generator:Design and Analysis. \newblock Computer Engineering and Science 2009;.
\bibitem[{Zeng(2010)Zeng, Guihua}]{zeng2010quantum} Zeng G. \newblock Quantum private communication. \newblock Springer Publishing Company, Incorporated; 2010.
\bibitem[{Zhang et~al.(2014)Zhang, Zhijun and Liu, Xingbing and Duan, Xintao}]{Zhang2014Algorithm} Zhang Z, Liu X, Duan X. \newblock Algorithm comparison on Gaussian random number generators. \newblock Journal of Henan Institute of Science and Technology 2014;.
\bibitem[{Zhou et~al.(2010)Zhou, Dong Mei and Zhang, Zhi Bin and Liu, Yu Hong}]{Zhou2010Study} Zhou DM, Zhang ZB, Liu YH. \newblock Study of m Sequence Generator SSRG and MSRG. \newblock Journal of Lanzhou Jiaotong University 2010;.
\end{thebibliography}

\end{document}